\documentclass[preprint,aps]{revtex4}
\begin{document}
\title{ON THE HYDROGEN ATOM VIA WIGNER-HEISENBERG ALGEBRA}
\author{R. de Lima Rodrigues\\
Unidade Acad\^emica de Educa\c{c}\~ao\\ Universidade Federal de
Campina Grande, Cuit\'e - PB, CEP 58.175-000- Brazil
\\ Centro Brasileiro de Pesquisas F\'\i sicas (CBPF)\\
Rua Dr. Xavier Sigaud, 150, CEP 22290-180, Rio de Janeiro, RJ, Brazil}

\begin{abstract}
We extend the usual Kustaanheimo-Stiefel $4D\rightarrow 3D$ mapping to study and discuss
a constrained super-Wigner oscillator in four dimensions. We show that
the physical hydrogen atom is the system that emerges
in the bosonic sector of the mapped super $3D$ system.

 \vspace{2cm} PACS numbers: 11.30.Pb,
03.65.Fd, 11.10.EF

\vspace{4cm} E-mail to RLR is rafael@df.ufcg.edu.br or
rafaelr@cbpf.br. J. Phys. A: Math. Theor. (2009) {\bf 42} 355213.

\end{abstract}

\maketitle

\newpage

\section{Introduction}

\paragraph*{}

The R-deformed Heisenberg or Wigner-Heisenberg (WH) algebraic technique
\cite{WH} which was super-realized for
quantum oscillators \cite{JR90,JR94,Mik00},
is related to the paraboson relations introduced by Green \cite{Green53}.

Let us now point out that the WH algebra is given by following
(anti-)commutation relations ($[A,B]_+\equiv
AB+BA$ and $[A,B]_-\equiv AB-BA):$

\begin{equation}
\label{RH}
H=\frac 12 [a^-, a^+]_+, \quad
[H, a^{\pm}]_-=\pm a^{\pm}, \quad
 [a^-, a^+]_-=1 + cR,
\end{equation}
where $c$ is a real constant associated to the Wigner parameter
\cite{JR90} and the $R$ operator satisfies

\begin{equation}
\label{Ra}
[R, a^{\pm}]_+=0, \quad R^2=1.
\end{equation}
Note that when $c=0$ we have the
standard Heisenberg algebra.

The generalized quantum condition given in Eq. (\ref{RH}) has been
found relevant in the context of integrable models \cite{Vasiliev91}.
Furthermore, this algebra was also used to solve the energy
eigenvalue and eigenfunctions
of the Calogero interaction, in the context of one-dimensional
many-body integrable systems, in terms of a new set of phase
space variables involving exchanged operators \cite{Poly92,macfa93}.
From this WH algebra a new kind of deformed calculus has been developed
\cite{Jing98,Matos01,R03}.

The WH algebra has been
considered for the three-dimensional non-canonical oscillator to
generate a representation of the orthosympletic Lie superalgebra
$osp(3/2)$ \cite{PS}, and recently Palev {\it et. al.} have
investigated the $3D$ Wigner oscillator under a discrete non-commutative
context \cite{palev03,palev06}. Also, the connection of the WH algebra with
the Lie superalgebra $s\ell(1|n)$  has been studied
in a detailed manner \cite{stoilova06}.

Recently,  the relevance of
relations (\ref{RH}) to
quantization in fractional dimension has been also discussed
\cite{Matos01b,LT04}
and the properties of Weyl-ordered polynomials in operators $P$ and $Q,$ in
fractional-dimensional quantum mechanics have been developed \cite{LT05}.

The Kustaanheimo-Stiefel mapping \cite{KS65} yields the Schr\"odinger equation for the hydrogen atom that   has been
exactly solved and well-studied in the literature. (See for example,
Chen \cite{chen80}, Cornish \cite{Corn84}, Chen and Kibler
\cite{CK85}, D'Hoker and Vinet \cite{HV85}.) Kostelecky, Nieto and
Truax \cite{KNT85} have studied in a detailed manner the relation of
the supersymmetric (SUSY) Coulombian problem
\cite{KN84,Amado88,Lange91,Tangerman93,bjd} in D-dimensions with
that of SUSY isotropic oscillators in D-dimensions in the radial
version. (See also Lahiri {\it et. al.} \cite{Lahi87}) For the
mapping with $3D$ radial oscillators, see also Bergmann and Frishman
\cite{BF65}, Cahill \cite{Cahill90} and J. - L. Chen {\it et. al.}
\cite{Chen00}.

The connection of the D-dimensional hydrogen
atom with the D-dimensional harmonic oscillator in terms of the $su(1,1)$ algebra has been
investigated by Gao-Jian Zeng
{\it et. al.} \cite{Gao94}. However, the correspondence mapping of a $4D$
isotropic constrained super Wigner oscillator (for super Wigner
oscillators see our previews work \cite{JR90,JR94}) with the corresponding
super system in $3D$ such that the usual $3D$ hydrogen atom emerges
in the $4D \rightarrow 3D$ mapping in the bosonic sector has not
been studied in the literature; the objectives of the present work are
to do such a mapping and to analyze in detail the consequences.
In this work, the stationary states
of the hydrogen atom are mapped onto the super-Wigner oscillator by using the
Kustaanheimo-Stiefel transformation.

This work is organized as follows. In Section II, we start by
summarizing the R-deformed Heisenberg algebra or Wigner-Heisenberg
algebraic technique for the  Wigner oscillator, based on the
super-realization of the WH algebra for simpler effective spectral
resolutions of general oscillator-related potentials, applied by
Jayaraman and Rodrigues, in Ref. \cite{JR90}. In Section III, we
illustrate how to construct the $4D \rightarrow 3D$ mapping in the
bosonic sector which offer's a simple resolution of the hydrogen
energy spectra and eigenfunctions. The conclusion is given in
Section IV.

\section{The Super Wigner Oscillator in 1D}

The Wigner oscillator ladder operators

\begin{equation}
\label{loa}
 a^{\pm} = \frac{1}{\sqrt {2}} (\pm i\hat {p}_x - \hat x)
\end{equation}
of the WH algebra may be written in terms of the
super-realization of the position and momentum operators viz., $\hat
x=x\Sigma_1$ and $\hat{p}_x=-i\Sigma_1\frac{d}{dx}+\frac{c}{2x}\Sigma_2,$
satisfy the general quantum rule 
$[\hat {x}, \hat {p}_x]_- = i (1 + cR),$ where $c=2(\ell +1).$ Thus, in this representation the
reflection operator becomes $R=\Sigma_3,$ where $\Sigma_3$ is the diagonal
Pauli matrix.

Thus, from the super-realized first order ladder operators  given by

\begin{equation}
\label{awh}
a^{\pm }(\ell +1) = {1\over \sqrt{2}}\left\{\pm {d\over dx} \pm
{(\ell +1)\over x}\Sigma_{3} - x\right\}\Sigma_{1}, \quad \ell > 0,
\end{equation}
the Wigner Hamiltonian becomes

\begin{equation}
\label{E6}
H(\ell +1) = {1\over 2}\left[a^{+}(\ell +1), a^{-}(\ell +1)\right]_{+}
\end{equation}
and the WH algebra ladder relations are readily obtained as

\begin{equation}
\label{E7}
\left[H(\ell +1), a^{\pm }(\ell +1)\right]_{-} =
\pm a^{\pm }(\ell +1).
\end{equation}
Equations (\ref{E6}) and (\ref{E7}) together with the commutation relation

\begin{equation}
\label{E8}
\left[a^{-}(\ell +1), a^{+}(\ell +1)\right]_{-}=
1 + 2(\ell +1)\Sigma_{3}
\end{equation}
constitute the super WH algebra.

Thus, the super Wigner oscillator Hamiltonian in terms of
the Pauli's matrices ($\Sigma_i,$ i=1,2,3) is given by

\begin{eqnarray}
\label{WH}
H(\ell +1) &=& {1\over 2}\left\{- {d^{2}\over dx^{2}} + x^{2} +
{1\over x^{2}} (\ell +1)[(\ell +1)\Sigma_{3} - 1]\Sigma_{3}\right\}
\nonumber\\
&=& \left(
\begin{array}{cc}
H_{-}(\ell )&0\\
0&H_{+}(\ell )=H_{-}(\ell +1)
\end{array}\right),
\end{eqnarray}
where the bosonic and fermionic sector Hamiltonians are respectively given by

\begin{equation}
\label{ich}
 H_{-}(\ell ) = {1\over 2}\left\{ - {d^{2}\over dx^{2}} +
x^{2} + {1\over x^2} \ell (\ell +1)\right\}
\end{equation}
and

\begin{equation}
H_{+}(\ell ) = {1\over 2} \left\{- {d^{2}\over dx^{2}} + x^{2} +
{1\over x^{2}} (\ell +1)(\ell +2)\right\}= H_{-}(\ell +1).
\end{equation}
Note that the bosonic sector is the Hamiltonian of the oscillator with barrier.

The super Wigner oscillator eigenfunctions that generate
the eigenspace associated with even(odd) $\Sigma_{3}$-parity for bosonic(fermionic)
quanta $n=2m(n=2m+1)$ are given by

\begin{equation}
\Psi_{n=2m}(x;\ell +1) = \left(\begin{array}{cc} 
\psi^{(m)}_-(x;\ell)
\\ 0
\end{array}\right),\quad
\Psi_{n=2m+1}(x; \ell +1) = \left(\begin{array}{cc} 0 \\
\psi^{(m)}_+(x; \ell)
\end{array}\right)
\end{equation}
and satisfy the following eigenvalue equation

\begin{eqnarray}
H(\ell +1)\Psi_{n}(x; \ell +1) &=& E_{n}\Psi_{n}(x; \ell +1)\nonumber\\
\Sigma_3\Psi_{n=2m}&=&\Psi_{n=2m}\nonumber\\ 
\Sigma_3\Psi_{n=2m+1}&=&-\Psi_{n=2m+1}
\end{eqnarray}
where the non-degenerate energy eigenvalues are obtained by the
repeated application of the raising operator on the ground eigenstate

\begin{equation}
\Psi_{n}(x; \ell +1) \propto (a^+(\ell + 1))^n\Psi_{0}(x; \ell +1)
\end{equation}
and are given by

\begin{equation}
E_{n} = \ell  + {3\over 2} + n, \quad n=0,1,2,\ldots .
\end{equation}
The ground state energy eigenfunction satisfies the following annihilation condition

\begin{equation}
\label{ca}
a^-(\ell +1)\Psi_{(0)}(x; \ell +1)=0,
\end{equation}
which using Eq. (\ref{awh}) result in
$$
\psi^{(0)}_-(x; \ell)=N_1x^{(\ell +1)}e^{-\frac{x^2}{2}}, \quad\psi^{(0)}_+(x; \ell)=N_2x^{-(\ell +1)}e^{-\frac{x^2}{2}}. 
$$
If we assume $\ell + 1 >0,$ only $\psi^{(0)}_-(x; \ell)$  meets the physical requirement
of vanishing at the origin and $\psi^{(0)}_+(x; \ell),$ which does not stand
this test, is discard by setting $N_2=0$ in (\ref{ca}). 

In this case,
the normalizable ground-state eigenfunction is given, up to a
normalization constant, by

\begin{equation}
\Psi_0(x; \ell +1)\propto
\left(\begin{array}{cc}
x^{(\ell +1)}e^{-\frac{1}{2}x^{2}}\\
0
\end{array}\right),
\label{ef}
\end{equation}
which has even $\Sigma_3$-parity, i.e.
$
\Sigma_3\Psi_0(x;\ell +1)=\Psi_0(x;\ell +1).
$

For the bosonic  and fermionic sector Hamiltonians the energy
eigenvectors  satisfy  the following equations

\begin{equation}
H_{\pm}(\ell )\psi_{\pm}^{(m)}(x; \ell) = E^{(m)}_{\pm}\psi_{\pm}^{(m)}(x; \ell),
\end{equation}
where the eigenvalues are exactly constructed via WH algebra ladder relations
and are given by 

\begin{equation}
E^{(m)}_{-}=
E_0 + 2m ,\qquad E^{(m)}_{+}=
E_0 + 2(m+1), \qquad m=0,1,2,\ldots,
\end{equation}
where $E_0$ is the energy of the Wigner oscillator ground state.
Note that the energy spectrum of a particle in a
potential given by bosonic sector Hamiltonian
is equally spaced like that of  the 3D
isotropic harmonic  oscillator, with difference of two quanta
between two levels.

Also, note that the operators $a^{\pm}(\ell+1)$ given in Eq. (\ref{awh}) together
with  $H(\ell+1), J_{\pm}=(a^{\pm}(\ell+1))^2$ satisfy an 
$osp(1\mid 2)$ superalgebra.

\section{The constrained Super Wigner Oscillator in $4D$}

\paragraph*{}

The usual isotropic oscillator in $4D$ has the following eigenvalue
equation for it's Hamiltonian $H^B_{\hbox{osc}}$, described by
(employing natural system of units $\hbar = m = 1$) time-independent
Schr\"odinger equation

\begin{equation}
\label{OB} H^{\hbox{B}}_{osc} \Psi^{\hbox{B}}_{osc}(y) =
E^{\hbox{B}}_{osc} \Psi^{\hbox{B}}_{osc}(y),
\end{equation}
with

\begin{equation}
\label{HOB}
H^{\hbox{B}}_{osc} = - \frac{1}{2} \nabla^{2}_{4} +
\frac{1}{2} s^{2},  \quad s^2 =
\Sigma^{4}_{i=1} y^{2}_{i},
\end{equation}

\begin{equation}
\label{L4D}
\nabla^{2}_{4} = \frac{\partial^2}{\partial y^2_{1}} +
\frac{\partial^2}{\partial y^2_{2}} + \frac{\partial^2}{\partial
y^2_{3}}+\frac{\partial^2}{\partial y^2_{4}} = \sum^{4}_{i=1}
\frac{\partial^2}{\partial y^2_{i}},
\end{equation}
where the superscript $B$ in $H^B_{\hbox{osc}}$ is in anticipation
of the Hamiltonian, with constraint to be defined, being implemented
in the bosonic sector of the super $4D$ Wigner system with unitary
frequency.  Changing to spherical coordinates in
4-space dimensions, allowing a factorization of the energy
eigenfunctions as a product of a radial eigenfunction and
spin-spherical harmonic.

In (\ref{L4D}), the coordinates $y_{i} (i = 1,2,3,4)$ in spherical coordinates in 4D are defined by \cite{chen80,HV85}

\begin{eqnarray}
\label{y123}
y_{1} &=& s \cos\left(\frac{\theta}{2}\right)
\cos\left(\frac{\varphi-\omega}{2}\right)\nonumber\\
y_{2} &=& s
\cos\left(\frac{\theta}{2}\right)
\sin\left(\frac{\varphi-\omega}{2}\right)\nonumber\\
y_{3} &=& s \sin\left(\frac{\theta}{2}\right)
\cos\left(\frac{\varphi+\omega}{2}\right)\nonumber\\
y_{4} &=& s \sin\left(\frac{\theta}{2}\right)
\sin\left(\frac{\varphi+\omega}{2}\right),
\end{eqnarray}
where $0\leq \theta\leq \pi, \quad
0\leq \varphi\leq 2\pi$ and $0\leq \omega \leq 4\pi$.

The mapping of the coordinates $y_i (i = 1, 2, 3, 4)$ in
$4D$ with the Cartesian coordinates $\rho_i (i = 1, 2, 3)$ in $3D$
is given by the Kustaanheimo-Stiefel transformation 

\begin{equation}
\label{rhoi}
\rho_{i} =\sum^{2}_{a,b=1} z^{*}_{a} \Gamma^{i}_{ab}
z_{b}, \quad {(i = 1,2,3)}
\end{equation}

\begin{equation}
\label{zi}
z_{1} = y_{1} + i y_{2}, \quad z_{2} = y_{3} + i y_{4},
\end{equation}
where the $\Gamma^{i}_{ab}$ are the elements of the usual Pauli matrices. 
If one defines $z_1$ and $z_2$ as in Eq. (\ref{zi}), 
$Z= \left(\begin{array}{cc} z_1 \\
z_2
\end{array}\right)$ is a two dimensional spinor of $SU(2)$ transforming
as $Z\rightarrow Z^{\prime}= UZ$ with $U$ a two-by-two matrix of $SU(2)$
and of course
$Z^{\dagger}Z$ is invariant. So the transformation (\ref{rhoi}) is very spinorial. Also, using the standard Euler angles parametrizing
$SU(2)$ as in transformations (\ref{y123}) and (\ref{zi}) one obtains

\begin{eqnarray}
\label{z12}
z_{1} &=& s\cos\left(\frac{\theta}{2}\right)
e^{\frac{i}{2}(\varphi-\omega)}\nonumber\\
z_{2} &=& s
\sin\left(\frac{\theta}{2}\right)
e^{\frac{i}{2}(\varphi+\omega)}.
\end{eqnarray}
Note that the angles in these equations are divided by two. 
However, in 3D, the angles are not divided by two, viz.,  
$\rho_3=\rho cos^2(\frac{\theta}{2})-\rho sin^2(\frac{\theta}{2})=
\rho cos\theta.$ Indeed, from (\ref{rhoi}) and (\ref{z12}), we obtain

\begin{equation}
\label{rhoi-1}
\rho_{1} = \rho \sin \theta \cos \varphi, \quad \rho_{2}
= \rho \sin \theta \sin \varphi, \quad \rho_{3} = \rho \cos \theta
\end{equation}
and also that

\begin{eqnarray}
\label{rho}
 \rho &=& \left\{\rho^{2}_{1} + \rho^{2}_{2} +
\rho^{2}_{3}\right\}^{\frac{1}{2}} =
\left\{(\rho_{1}+i\rho_{2})(\rho_{1} - i\rho_{2}) +
\rho^{2}_{3}\right\}^{\frac{1}{2}}\nonumber\\
{}&=&\left\{(2z^{*}_{1} z_{2}) (2 z_{1} z^{*}_{2}) +
(z^{*}_{1} z_{1} - z^{*}_{2}z_{2})^2\right\}^{\frac{1}{2}}\nonumber\\
{}&=&(z_1z^{*}_{1} + z_2z^{*}_{2}) = \sum^{4}_{i=1} y^{2}_{i}=s^2.
\end{eqnarray}
The complex form of the Kustaanheimo-Stiefel transformation was given 
by  Cornish \cite{Corn84}.

Thus, the expression for $H^{\hbox{B}}_{osc}$ in
(\ref{HOB}) can be written in the form 

\begin{eqnarray}
\label{Hosc}
H^{\hbox{B}}_{osc} &=&
-\frac{1}{2}\left(\frac{\partial^2}{\partial s^2}
+ \frac{3}{s} \frac{\partial}{\partial s}\right)  \nonumber\\
{}&-&\frac{2}{s^2} \left[\frac{1}{\sin\theta}
\frac{\partial}{\partial \theta} \sin \theta \frac{\partial}{\partial\theta}
+\frac{1}{sin^2 \theta} \frac{\partial^2}{\partial \varphi^2}
 +\frac{1}{sin^2\theta} \left(2cos\theta\frac{\partial}
{\partial\varphi} + \frac{\partial}{\partial \omega}\right)
\frac{\partial}{\partial\omega} \right] + \frac 12 s^2.
\end{eqnarray}
We obtain a constraint by projection (or "dimensional reduction") from
four to three dimensional.
Note that $\psi^{\hbox{B}}_{osc}$ is independent of $\omega$ provides the constraint condition

\begin{equation}
\label{ev}
\frac{\partial}{\partial \omega} \Psi^{\hbox{B}}_{osc}
(s,\theta,\varphi) = 0,
\end{equation}
imposed on $H^{\hbox{B}}_{osc}$, the expression for this
restricted Hamiltonian, which we continue to call as
$H^{\hbox{B}}_{osc}$, becomes

\begin{eqnarray}
\label{nHosc}
H^{\hbox{B}}_{osc} =
-\frac 12\left(\frac{\partial^2}{\partial s^2} + \frac{3}{s}
\frac{\partial}{\partial s}\right) - \frac{2}{s^2} \left[
\frac{1}{\sin \theta} \frac{\partial}{\partial \theta} \sin \theta
\frac{\partial}{\partial \theta} + \frac{1}{\sin^2\theta}
\frac{\partial^2}{\partial\varphi^2} \right] + \frac 12 s^2.
\end{eqnarray}

Identifying the expression in bracket in (\ref{nHosc}) with $L^2$, the square
of the orbital angular momentum operator in $3D,$ since we always have

\begin{equation}
\label{L2} L^2 = (\vec \sigma \cdot \vec L)(\vec
\sigma \cdot \vec L + 1),
\end{equation}
which is valid for any system, where $\sigma_{i} (i = 1,2,3)$ are the Pauli matrices representing
the spin $\frac{1}{2}$ degrees of freedom, we obtain for
$H^{\hbox{B}}_{osc}$ the final expression

\begin{equation}
\label{mHosc}
 H^{\hbox{B}}_{osc} = \frac{1}{2}
\left[-\left(\frac{\partial^2}{\partial s^2} + \frac{3} {s}
\frac{\partial}{\partial s}\right) + \frac{4}{s^2} (\vec
\sigma \cdot \vec L) (\vec \sigma \cdot \vec L +
1) + s^2 \right].
\end{equation}

Now, associating $H^{\hbox{B}}_{osc}$ with the bosonic sector of
the super Wigner system, $H_{\hbox{w}},$ subject to the same
constraint as in (\ref{ev}),  and following the analogy with the Section II
of construction of super Wigner systems, we first must solve
the Schr\"odinger equation

\begin{equation}
\label{Wae} H_{\hbox{w}} \Psi_{\hbox{w}} (s,\theta,\varphi) =
E_{\hbox{w}} \Psi_{\hbox{w}}(s,\theta,\varphi),
\end{equation}
where the explicit form of $H_{\hbox{w}}$ is given by

\begin{eqnarray}
\label{HW}
&&H_{\hbox{w}}(2\vec \sigma \cdot \vec L +
\frac{3}{2})=
\nonumber\\
{}&&\left(
\begin{array}{cc}
-\frac{1}{2}(\frac{\partial}{\partial s} + \frac{3}{2s})^2 + \frac 12 s^2 +
\frac{(2 \vec \sigma \cdot \vec L +
\frac{1}{2})(2 \vec \sigma
\cdot \vec L + \frac{3}{2})}{2s^2} & 0 \\
0 & -\frac{1}{2}{(\frac{\partial}{\partial s} +
\frac{3}{2s})}^2 + \frac 12 s^2 +
\frac {(2 \vec \sigma \cdot \vec L + \frac{3}{2})(2 \vec
\sigma \cdot \vec L + \frac{5}{2})}{2s^2}
\end{array}\right).
\end{eqnarray}

Using the operator technique in references \cite{JR90,JR94},
we begin with the
following super-realized mutually adjoint operators

\begin{eqnarray}
\label{a+-}
a^\pm_{\hbox{w}}\equiv a^\pm (2 \vec \sigma \cdot \vec L +
\frac{3}{2}) = \frac{1}{\sqrt{2}} \left[\pm
\left(\frac{\partial}{\partial s} + \frac{3}{2s}\right) \Sigma_{1}
\mp \frac{1}{s} (2 \vec \sigma \cdot \vec L +
\frac{3}{2}) \Sigma_{1} \Sigma_{3} - \Sigma_{1} s\right],
\end{eqnarray}
where $\vec{\Sigma}_i (i = 1, 2, 3)$ constitute a set of Pauli
matrices that provide the fermionic coordinates commuting with the
similar Pauli set $\sigma_i (i = 1, 2, 3)$ already introduced
representing the spin $\frac{1}{2}$ degrees of freedom.

It is checked, after some algebra, that $a^+$ and $a^-$ of (\ref{a+-})
are
indeed the raising and lowering operators for the spectra of the super
Wigner Hamiltonian $H_{\hbox{w}}$ and they satisfy the following
(anti-)commutation relations of the WH algebra:

\begin{eqnarray}
\label{nWH}
&&H_{\hbox{w}} = \frac{1}{2} [a^{-}_{\hbox{w}},a^{+}_{\hbox{w}}]_+ \nonumber\\
{}&&= a^{+}_{\hbox{w}} a^{-}_{\hbox{w}} + \frac{1}{2} \left[1 +
2(2\vec\sigma\cdot\vec L + \frac{3}{2})
\Sigma_{3}\right] \nonumber\\
{}&&= a^{-}_{\hbox{w}} a^{+}_{\hbox{w}} - \frac{1}{2} \left[1 +
2(2\vec\sigma\cdot\vec L + \frac{3}{2})
\Sigma_{3}\right]
\end{eqnarray}

\begin{equation}
\label{aWH}
[H_{\hbox{w}}, a^{\pm}_{\hbox{w}}]_{-} = \pm a^{\pm}_{\hbox{w}}
\end{equation}

\begin{equation}
\label{rcg} [a^{-}_{\hbox{w}},  a^{+}_{\hbox{w}}]_{-} = 1 +
2(2\vec\sigma\cdot\vec L + \frac{3}{2}) \Sigma_{3},
\end{equation}

\begin{equation}
\label{s3} [\Sigma_{3}, a^{\pm}_{\hbox{w}}]_{+} = 0 \Rightarrow [\Sigma_{3},
H_{\hbox{w}}]_{-} = 0.
\end{equation}

Since  the operator $\left(2 \vec \sigma
\cdot\vec L + \frac{3}{2}\right)$ commutes with the basic
elements $a^{\pm}, \Sigma_3$ and $H_{\hbox{w}}$ of the WH algebra
(\ref{nWH}), (\ref{aWH}) and (\ref{rcg})
it can be replaced by its eigenvalues $(2\ell +
\frac{3}{2})$ and $-(2\ell + \frac{5}{2})$ while acting on the
respective eigenspace in the from

\begin{equation}
\Psi_{\hbox{osc}} (s, \theta, \varphi) =\left(\begin{array}{c}
\Psi^{\hbox{B}}_{\hbox{osc}} (s, \theta, \varphi) \\
\Psi^{\hbox{F}}_{\hbox{osc}} (s, \theta, \varphi)\end{array}\right)
= \left(\begin{array}{c} R^{\hbox{B}}_{\hbox{osc}} (s) \\
R^{\hbox{F}}_{\hbox{osc}} (s) \end{array}\right) y_{\pm} (\theta,
\varphi)
\end{equation}
in the notation where $y_{\pm}(\theta, \varphi)$ are the
spin-spherical harmonics \cite{G65,MV78},

\begin{eqnarray}
y_{+}(\theta,\varphi)&=&y_{\ell\frac{1}{2};j=\ell+\frac{1}{2},m_{j}}(\theta,\varphi)
\nonumber\\
y_{-}(\theta,\varphi)&=&y_{\ell+1\frac{1}{2};j=(\ell+1)-\frac{1}{2},m_{j}}(\theta,\varphi)
\label{hs}
\end{eqnarray}
so that, we obtain:
$(\vec{\sigma} \cdot\vec{L} + 1) y_{\pm} = \pm(\ell +1) y_{\pm}, \quad (2\vec{\sigma}\cdot \vec{L} + \frac{3}{2})y_+ =
(2\ell + \frac{3}{2})y_+$ and 
$(2\vec{\sigma}\cdot \vec{L} + \frac{3}{2})y_- =
-[2(\ell + 1) + \frac{1}{2}]y_-.$ 
Note that on these subspaces the 3D
WH algebra is reduced to a formal 1D radial form with
$H_{\hbox{w}} (2\vec{\sigma} \cdot \vec{L} +
\frac{3}{2})$ acquiring respectively the forms $H_{\hbox{w}} (2\ell
+ \frac{3}{2})$ and

\begin{equation}
\label{NWH}
H_{\hbox{w}} \left(-2\ell - \frac{5}{2}\right)
= \Sigma_{1} H_{\hbox{w}} \left(2 \ell + \frac{3}{2}\right) \Sigma_{1}.
\end{equation}

Thus, the positive finite form of $H_{\hbox{w}}$ in (\ref{nWH})
together with the ladder relations (\ref{aWH})
and the form (\ref{rcg}) leads to the direct determination
of the state energies and the corresponding Wigner ground state wave
functions by the simple application of the annihilation conditions

\begin{equation}
a^- (2\ell + \frac{3}{2}) \left(\begin{array}{c}
R^{B^{(0)}}_{\hbox{osc}} (s) \\
 R^{F^{(0)}}_{\hbox{osc}}
(s)\end{array}\right) = 0.
\end{equation}
Then, the complete energy spectrum for $H_{\hbox{w}}$ and the whole set of
energy eigenfunctions $\Psi^{(n)}_{\hbox{osc}} (s, \theta ,\varphi) (n =
2m, 2m + 1, m = 0, 2 \cdots )$ follows from the step up operation
provided by $a^+(2\ell + \frac{3}{2})$ acting on the ground state,
which are also simultaneous
eigenfunctions of the fermion number operator $N = \frac{1}{2} (1 -
\Sigma_3).$ We obtain for the bosonic sector Hamiltonian
$H^{B}_{\hbox{osc}}$ with fermion number $n_f = 0$ and even orbital
angular momentum $\ell_{4} = 2\ell, (\ell = 0, 1, 2\ldots ),$
the complete energy spectrum and
eigenfunctions given by

\begin{equation}
\label{EB}
 \left[E^{B}_{osc}\right]^{(m)}_{\ell_{4} = 2\ell} =
2\ell + 2 + 2m , \quad (m = 0,1,2, \ldots),
\end{equation}

\begin{equation}
\label{psiB}
\left[\Psi^{B}_{osc}
(s,\theta,\varphi)\right]^{(m)}_{\ell_{4}=2\ell} \propto s^{2\ell}
\exp{\left(- \frac{1}{2} s^2\right)} L^{(2\ell+1)}_{m}(s^2)
\left\{
\begin{array}{l}
y_{+} (\theta, \varphi)
\\
y_{-}(\theta, \varphi)
\end{array}
\right.
\end{equation}
where $L^\alpha_m (s^2)$ are generalized Laguerre polynomials \cite{JR90}.
Now, to relate
the mapping of the 4D super Wigner system given by (\ref{WH})
with the
corresponding  system in 3D, we make use of the substitution of $s^2
= \rho$, Eq. (\ref{ev})
and the following substitutions

\begin{equation}
\label{d2s} \frac{\partial}{\partial s} = 2 \sqrt{\rho}
\frac{\partial}{\partial \rho}, \quad \frac{\partial^2}{\partial
s^2} = 4\rho \frac{\partial^2}{\partial\rho^2} + 2
\frac{\partial}{\partial \rho},
\end{equation}
in (\ref{HW}) and divide the eigenvalue equation for $H_{\hbox{w}}$
in (\ref{Wae}) by $4s^2 = 4\rho,$ obtaining

\begin{eqnarray}
\label{} &&\left(
\begin{array}{cc}
-\frac{1}{2} \left(\frac{\partial^2}{\partial \rho^2} +
\frac{2}{\rho}
\frac{\partial}{\partial \rho}\right) - \frac{1}{2}
\left[- \frac{1}{4} -  \frac{\vec \sigma \cdot
\vec L (\vec \sigma
\cdot \vec L + 1)}
{\rho^2}\right] & 0 \\
0 & -\frac{1}{2} \left(\frac{\partial^2}{\partial \rho^2} +
\frac{2}{\rho}
\frac{\partial}{\partial \rho}\right) - \frac{1}{2}
\left[- \frac{1}{4} -  \frac{(\vec \sigma \cdot
\vec L +\frac{1}{2})(\vec \sigma
\cdot \vec L + \frac{3}{2})}
{\rho^2}\right]
\end{array}\right)
\left(
\begin{array}{c}
\Psi^{B} \\
\Psi^{F}
\end{array}\right)\nonumber\\
{}&&=
\frac{1}{4\rho} E_{\hbox{w}}
\left(
\begin{array}{c}
\Psi^{B} \\
\Psi^{F}
\end{array}\right).
\end{eqnarray}

    The bosonic sector of the above eigenvalue equation can immediately be
identified with the eigenvalue equation for the Hamiltonian of the
3D Hydrogen-like atom expressed in the equivalent form given by

\begin{equation}
\label{eqh} \left\{-\frac{1}{2} \left(\frac{\partial^2}{\partial
\rho^2} + \frac{2}{\rho} \frac{\partial}{\partial\rho}\right) -
\frac{1}{2} \left [- \frac{1}{4} - \frac{\vec \sigma \cdot
\vec L (\vec \sigma \cdot \vec L + 1)}
{\rho^2}\right]\right\} \psi(\rho,\theta, \varphi) =
\frac{\lambda}{2\rho} \psi(\rho,\theta, \varphi),
\end{equation}
where $\Psi^{B}=\psi(\rho,\theta, \varphi)$ and the connection between the
dimensionless and dimensionfull eigenvalues, respectively, $\lambda$ and
$E_a$ with $e=1=m=\hbar$ is given by \cite{MV78}

\begin{equation}
\label{lamb}
\lambda = \frac{Z}{\sqrt{-2E_a}}, \quad \rho = \alpha
r, \quad \alpha = \sqrt{-8E_a},
\end{equation}
where $E_a$ is the energy of the electron Hydrogen-like atom, $(r,
\theta, \varphi)$ stand for the spherical polar coordinates of the
position vector $\vec{r} = (x_1, x_2, x_3)$ of the electron in
relative to the nucleons of charge $Z$
together with $s^2 = \rho.$
We see then  from equations (\ref{EB}), (\ref{psiB}), (\ref{eqh}) and (\ref{lamb})
that the complete energy spectrum and
eigenfunctions for the Hydrogen-like atom given by

\begin{equation}
\label{nEB} \frac{\lambda}{2} = \frac{E^{B}_{osc}}{4} \Rightarrow
 [E_a]^{(m)}_{\ell} = [E_a]^{(N)} =
-\frac{Z^2}{2 N^2}, \quad (N = 1, 2,\ldots)
\end{equation}
and

\begin{equation}
\label{ER} \left[\psi{(\rho,\theta,\varphi)}\right]^{(m)}_{\ell;
\j,m_j} \propto\rho^{\ell} \exp{(- \frac{\rho}{2})}
L^{(2\ell+1)}_{m} {(\rho)} \left\{
\begin{array}{l}
y_{+}(\theta,\varphi)
\\
y_{-}(\theta,\varphi)
\end{array}\right.
\end{equation}
where $E^{B}_{osc}$ is given by Eq. (\ref{EB}).

Here, $N = \ell + m + 1$ $(\ell = 0, 1, 2, \cdots, N-1; m= 0, 1,
2,\cdots)$ is the principal quantum number. Kostelecky and Nieto
shown that the supersymmetry in non-relativistic quantum mechanics
may be realized in atomic systems \cite{KN84}. 

\section{CONCLUSION}
\label{sec:level5}

In this work, we  have deduced the energy eigenvalues and
eigenfunctions of the hydrogen atom via Wigner-Heisenberg (WH)
algebra in non-relativistic quantum mechanics. Indeed, from the
ladder operators for the 4-dimensional ($4D$) super Wigner system,
ladder operators for the mapped super $3D$ system, and hence for
hydrogen-like atom in bosonic sector, are deduced. The
complete spectrum for the hydrogen atom is found with considerable
simplicity. Therefore, the solutions of the time-independent
Schr\"odinger equation for the hydrogen atom were mapped onto the
super Wigner harmonic oscillator in $4D$ by using the
Kustaanheimo-Stiefel transformation.


\newpage

\centerline{\bf Acknowledgments}

RLR would like to acknowledge  CBPF for hospitality. The author
also like to acknowledge CES-UFCG of Cuit\'e-PB, Brazil. This
research was supported in part by CNPq (Brazilian Research Agency).
This work was initiated in collaboration with Jambunatha Jayaraman
(In memory), whose advises and encouragement were fundamental.



\end{document}